\documentclass[a4paper,11pt]{article}
\pdfoutput=1
\usepackage{jheppub} % for details on the use of the package, please
\usepackage{graphicx,color,rotating}
\usepackage{hyperref}
\usepackage{epsfig,color}
\usepackage{slashed}
\usepackage{amsfonts}
\usepackage{ulem}
\newcommand{\lsim}{\raisebox{-0.13cm}{~\shortstack{$<$ \\[-0.07cm] $\sim$}}~}

\usepackage{mathrsfs}
\usepackage{amsmath}
\usepackage{amsfonts}
\usepackage{amssymb}
\usepackage{tabularx}
\usepackage{graphicx}
\usepackage{adjustbox}
\usepackage{tabularx}

\begin{document}
\title{Boosted dark matter from primordial black holes produced in a first-order phase transition}

\renewcommand{\thefootnote}{\arabic{footnote}}

\author{
Danny Marfatia$^{1}$ and
Po-Yan Tseng$^{2}$}
\affiliation{
%$^1$ Department of Physics, University of Illinois at Chicago,
%Illinois 60607 USA \\
%%
%$^2$ Physics Division, National Center for Theoretical Sciences,
%Hsinchu, Taiwan \\
%%
$^1$ Department of Physics \& Astronomy, University of Hawaii at Manoa,
2505 Correa Rd., Honolulu, HI 96822, USA \\
$^2$ Department of Physics, National Tsing Hua University,
101 Kuang-Fu Rd., Hsinchu 300044, Taiwan R.O.C. \\
}
%\pacs{14.80.Bn.,14.80.Da,14.80.Ec}
\date{\today}

\abstract{During a cosmological first-order phase transition in a dark sector, fermion dark matter particles $\chi$ can form macroscopic Fermi balls that collapse 
to primordial black holes (PBHs) under certain conditions. The evaporation of the PBHs produces a boosted $\chi$ flux, which may be detectable
 if $\chi$ couples to visible matter. We consider the interaction of $\chi$ with electrons, and calculate signals of the dark matter flux in 
 the XENON1T, XENONnT, Super-Kamiokande and Hyper-Kamiokande experiments. A correlated gravitational wave signal from the phase transition
 can be observed at THEIA and $\mu$Ares. An amount of dark radiation measurable by CMB-S4 is an epiphenomenon of the phase transition.
}

\maketitle

\section{Introduction}

Dark matter (DM) direct detection experiments have sought a signal
of weakly interactive massive particles of ${\cal O}(100)$~GeV mass for decades. 
In recent years, the priority of these searches has shifted to GeV and sub-GeV DM. 
Because typical detectors have energy thresholds of a keV, they are unable to detect sub-GeV galactic halo DM particles with velocity $v/c\sim 10^{-3}$.
However, if the DM particles are energetically boosted, this hurdle is overcome.
Mechanisms for boosted DM include scattering by cosmic-rays~\cite{Bringmann:2018cvk,Ema:2018bih,Cappiello:2018hsu}, 
scattering by stellar and supernova neutrinos~\cite{Jho:2021rmn,Das:2021lcr}, and 
PBH evaporation to DM particles~\cite{Fujita:2014hha,Lennon:2017tqq,Morrison:2018xla,Masina:2020xhk,Cheek:2021odj}.
We focus on the latter in this work.

We consider PBH formation from the collapse of nontopological solitons, called Fermi balls, produced in a cosmological first-order phase transition (FOPT) in a dark sector.
The physical picture is as follows. Dark fermions $\chi$ get trapped in the false vacuum if they are heavier in the true vacuum than the false vacuum by at least the 
critical temperature of the FOPT. 
Then, as bubbles of the true vacuum fill the space 
the dark fermions trapped in the false vacuum aggregate to form Fermi balls (FBs)~\cite{Hong:2020est}. 
The Yukawa force responsible for making $\chi$ heavier in the true vacuum causes the FBs to collapse to PBHs once the separation distance between $\chi$ in the FB
becomes smaller than the range of the force~\cite{Kawana:2021tde}.

In Refs.~\cite{Marfatia:2021twj,Marfatia:2021hcp}, we implemented the FB scenario with only gravitational interactions between the dark and visible sectors, 
and a finite-temperature quartic effective potential for the FOPT.
As shown in Ref.~\cite{Marfatia:2021hcp}, a FOPT  with vacuum energy, \mbox{$0.1\lesssim B^{1/4}/{\rm MeV} \lesssim 10^3$}, produces PBHs of mass $10^{-20}\lesssim M_{\rm PBH}/M_\odot \lesssim 10^{-16}$ with Hawking temperatures, $500 \gtrsim T_{\rm PBH}/{\rm MeV} \gtrsim 0.05$. The extragalactic $\mathcal{O}(0.1-100)$~MeV gamma-ray background 
from PBH evaporation can be correlated with the stochastic gravitational wave (GW) background from the FOPT that peaks in the $10^{-7}~{\rm Hz}-10^{-4}~{\rm Hz}$ range. Note that no couplings to the visible sector are postulated to predict these correlated signals.

A PBH will emit any particle in the visible and dark sectors with mass below the Hawking temperature. 
The $\chi$ emitted in PBH evaporation will be relativistic if its mass is much smaller than $T_{\rm PBH}$, as is often the case~\cite{Marfatia:2021hcp}.
Now permitting nongravitational interactions with visible matter, we suppose that $\chi$ scatters with electrons with a cross section $\sigma_{\chi e}$. Then, the flux of 
boosted $\chi$ will produce a signal above the threshold energies 
of the XENON1T, XENONnT, Super-Kamiokande (SK) and Hyper-Kamiokande (HK)~\cite{XENON:2020rca,Aprile:2022vux,Super-Kamiokande:2017dch,Ema:2018bih,Hyper-Kamiokande:2018ofw} detectors. In fact, $\chi$ does not have to be relativistic to produce electron scattering signals in these detectors,
it just has to be boosted to velocities well in excess of $10^{-3}c$.
For the xenon detectors, we compute the differential $\chi-e$ ionization cross section by taking into account the bound electron wavefunctions in the Xenon atom. We only 
consider $ \sigma_{\chi e} < 10^{-31}~{\rm cm^2}$ so that attenuation in the Earth and atmosphere can be neglected~\cite{Ema:2018bih,Cappiello:2019qsw}. 
We also take care that the  $\chi-e$ interaction does not disrupt FB formation/stability and the FOPT.

The structure of the paper is as follows. In section~\ref{sec:FOPT}, we briefly describe the mechanism for PBH formation from a FOPT including the effect of $\chi-e$ interactions.  In section~\ref{sec:PBH_evaporation}, we estimate the $\chi$ flux from PBH evaporation. In section~\ref{sec:event_rate}, we calculate the expected DM event rates. We explore the parameter space that gives correlated DM and GW signals in section~\ref{sec:scan}, and summarize our results in section~\ref{sec:summary}.

\section{PBH formation}
\label{sec:FOPT}

We briefly describe how PBHs form during a FOPT, partially with a view to defining our notation, and refer the reader to Refs.~\cite{Marfatia:2021twj,Marfatia:2021hcp} for elaboration. 

We consider a dark sector described by the Lagrangian~\cite{Marfatia:2021hcp},
\begin{eqnarray}
\label{eq:Lagrangian}
\mathcal{L}\supset \bar{\chi}( i \slashed{\partial}-m) \chi  -g_\chi \phi \bar{\chi} \chi -V_{\rm eff}(\phi,T) \,,
\end{eqnarray}
 where the quartic thermal effective potential~\cite{Dine:1992wr,Adams:1993zs},
\begin{eqnarray}
\label{eq:quartic_potential}
V_{\rm eff}(\phi,T)= D(T^2-T^2_0)\phi^2-(AT+C)\phi^3+\frac{\lambda}{4}\phi^4\,,
\end{eqnarray}
 triggers a FOPT in the early Universe. Here, $2D(T^2-T^2_0)$ is the square of the thermal mass, with $T_0$ the 
 temperature at which the minimum at the origin is a local maximum.
At the critical temperature $T_c$, the Universe has two vacua corresponding
to degenerate local minima of $V_{\rm eff}$.
As the temperature falls below $T_c$, the FOPT begins and the Universe starts decaying
from the false vacuum ($\langle \phi \rangle=0$) 
to the true vacuum ($\langle \phi \rangle = v_\phi(T)$).
The characteristic energy scale of the FOPT is defined at zero temperature by the
vacuum energy density at the global minimum 
$\tilde{\phi}$, 
 $B\equiv - V_{\rm eff}(\tilde{\phi},0)$~\cite{Marfatia:2021twj}. 
 We neglect the logarithmic temperature dependence of $\lambda$, and treat it as a free parameter.
  Then, the free parameters that determine the potential are $\lambda, A, B, C$ and $D$, where we have traded
   $T_0$ for $B$. We define
 the temperature of the phase transition $T_\star$ as the temperature when the fractional volume of space in
 the false vacuum is $1/e$; this is also the percolation temperature. 
 We allow the dark sector to not be fully thermalized, so that the total radiation density is $\rho(T) = \frac{\pi^2}{30} g_\ast T_{\rm SM}^4$, where $g_\ast = g^{\rm SM}_\ast +g^{\rm D}_\ast\left(T/T_{\rm SM}\right)^4$ is the total number of relativistic degrees of freedom when the SM sector is at temperature $T_{\rm SM}$. 
 At all relevant times, $g^{\rm D}_\ast =4.5$. The latent heat converted to dark radiation during the FOPT heats the dark sector from $T_\star$ to $T_f$~\cite{Marfatia:2021twj}.
 To estimate the bubble wall velocity $v_w$, we only include the radiation energy density in the dark sector at $T_\star$. Since the amplitude
of the GW depends linearly on $v_w$, including the radiation energy density in the visible sector does not significantly change our results.

The Yukawa term in Eq.~(\ref{eq:Lagrangian})
contributes an additional mass $g_\chi v_\phi$ to $\chi$ in the true vacuum. 
If $T_c \ll g_\chi v_\phi$,
a $\chi$ located in the false vacuum is unable to enter the true vacuum
because it does not have enough thermal kinetic energy to provide the extra mass.
Consequently, $\chi$'s get trapped in the false vacuum and form FBs
once adequate degeneracy pressure develops to balance the vacuum pressure.
This presupposes a nonzero asymmetry 
$\eta_\chi \equiv (n_\chi-n_{\bar{\chi}})/s$ 
in the number densities (normalized to the entropy density $s$) so that $\bar{\chi}\chi \to \phi \phi$ annihilation is not complete~\cite{Hong:2020est}.
Calculations of the fraction of $\chi$ trapped in the false vacuum depend on the thickness and profile of the bubble wall, and are highly uncertain~\cite{Azatov:2021ifm}. Our estimates give a trapping fraction of ${\cal{O}}(0.5)$, which for the parameters and scales of our interest yields a negligible $\chi$ relic abundance~\cite{Hong:2020est}. We make the simplifying assumption that all the $\chi$'s are trapped in the false vacuum. This amounts to a rescaling of the value
of $\eta_\chi$.
Then, the total number of $\chi$'s in a FB is $Q_{\rm FB}=\eta_\chi (s/n_{\rm FB})$, where $n_{\rm FB}$ is the number density of FBs
produced in the FOPT.  We denote the mass and radius of a FB by $M_{\rm FB}$ and  $R_{\rm FB}$, respectively. 
The following conditions must be met for FBs to not decay to $\chi$ or fission to lighter FBs~\cite{Huang:2022him}:
\begin{eqnarray}
\frac{dM_{\rm FB}}{dQ_{\rm FB}} < m_\chi(T)\,,
~~{\rm and}~~\frac{d^2M_{\rm FB}}{dQ_{\rm FB}^2}< 0\,,
\label{stab}
\end{eqnarray}
where $m_\chi(T) =m+g_\chi v_\phi(T)$ is the $\chi$ mass in the true vacuum.
We checked that $\chi$-electron scattering does not disrupt FB formation or lead to FB instability as follows. Since $M_{\rm FB} \gg T_{\rm {SM\star}}$, stability against fission as dictated by the second relation in Eq.~(\ref{stab}), is unaffected.  If an electron scatters with $\chi$ inside a FB, the $\chi$ may be ejected from the FB unless its mean free path is short enough that
multiple scattering with other $\chi$ in the FB slows it down sufficiently. We find that the $\chi\chi$ scattering cross section  via $\phi$ exchange $\sigma_{\chi\chi}$ is larger than 
$\sigma_{\chi e}$ (assumed to be $<10^{-31}~{\rm cm^2}$) by many orders of magnitude. The mean free path $\ell_\chi \equiv (n_\chi \sigma_{\chi\chi})^{-1}$ turns out to be much smaller 
than $R_{\rm FB}$ so the scattered $\chi$ remains confined to the FB; see Table~\ref{tab:BP}.

To prevent $\chi$-electron scattering from disrupting FB formation, 
we require $m_\chi(T_\star)-m$ to be large enough that an electron
cannot excite $\chi$ from the false vacuum (in which FBs form) to the true vacuum. Specifically, we require $m_\chi(T_\star)-m> 2 T_{\rm {SM\star}}$.
Also, we assume that the rate for $\bar{\chi}\chi \to e^+e^-$ is negligible compared to $\bar{\chi}\chi \to \phi \phi$. This is easily accommodated because we have not prescribed a
Lagrangian for the $\chi$-electron interaction, and the relationship between the $\bar\chi\chi \to e^+e^-$ and $\chi-e$ scattering cross sections is model-dependent. 
For example, for the $\bar\chi \chi \bar e e$ operator, which leads to spin-independent scattering and $p$-wave annihilation~\cite{Kumar:2013iva}, we find the
 $\bar{\chi}\chi \to e^+e^-$  cross section to be ${\cal{O}}(\sigma_{\chi e})$, which is negligible compared to the $\bar{\chi}\chi \to \phi \phi$ cross section; 
 see Table~\ref{tab:BP}.

\begin{table}[t]
\caption{\small \label{tab:BP}
Benchmark points with $A=0.1$. 
%$T_f$ is the temperature of the dark sector after the FOPT.  
The strength of the FOPT is defined by $\alpha$, the ratio of the latent heat released and the
radiation energy density at the time of the FOPT, and $\beta^{-1}$ is its duration.
For the background only hypothesis, $\chi^2_{\rm bkgd}|_{\rm XENON1T}=167.1$ and $\chi^2_{\rm bkgd}|_{\rm XENONnT}=19.95$. The numbers in boldface correspond to a $>2\sigma$ signal.}
%\begin{ruledtabular}
\begin{adjustbox}{width=\textwidth}
\begin{tabular}{c|cc|cc|cc}
\hline
\hline
     & {\bf BP-1} & {\bf BP-2} & {\bf BP-3} & {\bf BP-4} & {\bf BP-5} & {\bf BP-6}  \\
\hline
\hline
$\lambda$     
             & 0.170 & 0.080 & 0.179 & 0.184 & 0.193 & 0.163   \\
$B^{1/4}/{\rm MeV}$     
             & 3.035 & 14.99 & 3.562 & 9.055 & 0.184 & 0.528  \\
$C/{\rm MeV}$     
             & 0.354 & 0.593 & 0.450 & 1.556 & 0.029 & 0.046 \\
$D$          & 0.408 & 0.383 & 0.371 & 0.560 & 0.350 & 0.706 \\
$g_\chi$     & 1.697 & 1.275 & 1.773 & 1.499 & 1.648 & 1.736 \\
$\eta_\chi$     
            & $5.29\times 10^{-15}$ 
            & $3.10\times 10^{-14}$ & $3.81\times 10^{-15}$
            & $7.84\times 10^{-16}$
            & $2.08\times 10^{-18}$ & $5.94\times 10^{-17}$ \\
$m/{\rm MeV}$ & 1.578 & 0.285 & 0.198 & 0.789 & 0.005 & 0.045 \\
$\sigma_{\chi e}/{\rm cm^2}$ 
            & $9.58\times 10^{-32}$ 
            & $6.49\times 10^{-32}$ & $9.91\times 10^{-32}$
            & $9.76\times 10^{-32}$
            & $1.60\times 10^{-34}$ & $8.50\times 10^{-32}$\\
%$r_T$       & 0.456 & 0.601 & 0.340 & 0.515 & 0.557 & 0.494 \\
\hline
$T_{{\rm SM}\star}/{\rm MeV}$     
            & 6.436 & 37.04 & 8.984 & 19.47 & 0.412 & 0.956 \\
$T_\star/{\rm MeV}$     
            & 2.756 & 13.84 & 3.455 & 5.918 & 0.184 & 0.352 \\
$T_f/{\rm MeV}$     
            & 2.961 & 15.99 & 3.677 & 7.907 & 0.198 & 0.383 \\
$T_{\phi}/{\rm MeV}$     
            & 2.694 & 13.79 & 3.396 & 5.019 & 0.165 & 0.350 \\
$T_0/{\rm MeV}$     
            & 1.729 & 10.08 & 2.043 & 0.958 & 0.084 & 0.280 \\
\hline
$\alpha$   
            & $5.06\times 10^{-2}$ 
            & $3.34\times 10^{-2}$ & $2.65\times 10^{-2}$ 
            & $4.03\times 10^{-2}$
            & $5.07\times 10^{-2}$ & $6.40\times 10^{-2}$ \\
$\beta/H_\star$    
            & $2.62\times 10^3$
            & $1.90\times 10^3$ & $2.52\times 10^3$
            & $1.15\times 10^3$
            & $2.22\times 10^3$  & $4.85\times 10^3$ \\
$v_w$                  
            & 0.978 & 0.983 & 0.973 & 0.993 & 0.971 & 0.990 \\
\hline
$dM_{\rm FB}/dQ_{\rm FB}/{\rm MeV}$   
            & 12.30 & 66.01 & 14.91 & 32.54 & 0.803 & 1.567 \\
$m_\chi(T_\star)/{\rm MeV}$ & 16.00 & 75.96 & 18.19 & 42.05 & 0.917 & 2.027 \\
$Q_{\rm FB}$
            & $1.48\times 10^{41}$ 
            & $1.55\times 10^{40}$ & $4.62\times 10^{40}$
            & $1.38\times 10^{40}$
            & $4.17\times 10^{41}$ & $6.76\times 10^{40}$ \\
$R_{\rm FB}$  [cm]    
            & 209 & 17.3 & 118 & 31.81 & 4525 & 1247 \\
$\sigma_{\chi \chi}$  [${\rm cm^2}$]   
            & $6.07\times 10^{-23}$ 
            & $3.05\times 10^{-24}$ & $3.25\times 10^{-23}$
            & $7.96\times 10^{-24}$
            & $8.79\times 10^{-21}$ & $3.47\times 10^{-21}$ \\
$n_\chi$ [${\rm cm^{-3}}$]    
            & $3.86\times 10^{33}$
            & $7.12\times 10^{35}$ & $6.71\times 10^{33}$
            & $1.02\times 10^{35}$
            & $1.07\times 10^{30}$ & $8.33\times 10^{30}$ \\
$\ell_\chi=1/n_\chi \sigma_{\chi \chi}$ [cm]     
            & $4.27\times 10^{-12}$
            & $4.60\times 10^{-13}$ & $4.58\times 10^{-12}$
            & $1.23\times 10^{-12}$
            & $1.06\times 10^{-10}$ & $3.46\times 10^{-11}$ \\
\hline
$M_{\rm PBH}/M_{\odot}$ 
            & $1.63\times 10^{-18}$ 
            & $9.15\times 10^{-19}$ & $6.16\times 10^{-19}$ 
            & $4.14\times 10^{-19}$
            & $2.97\times 10^{-19}$ & $9.56\times 10^{-20}$ \\
$\beta'$ 
            & $5.13\times 10^{-25}$ 
            & $1.21\times 10^{-23}$ & $2.75\times 10^{-25}$ 
            & $1.04\times 10^{-25}$
            & $5.56\times 10^{-30}$ & $1.79\times 10^{-28}$\\
$\Omega_{\rm PBH} h^2$
            & $1.90\times 10^{-8}$ 
            & $5.89\times 10^{-7}$ & $1.64\times 10^{-8}$
            & $7.57\times 10^{-9}$
            & $4.87\times 10^{-13}$ & $2.70\times 10^{-11}$ \\
$f_{\rm PBH} \times \sigma_{\chi e}/{\rm cm^2}$
            & $1.52\times 10^{-38}$ 
            & $3.19\times 10^{-37}$ & $1.35\times 10^{-38}$
            & $6.16\times 10^{-39}$
            & $6.48\times 10^{-46}$ & $1.91\times 10^{-41}$  \\
$\Delta N_{\rm eff}$
            & 0.406 & 0.234 & 0.223 & 0.268 & 0.451 & 0.452   \\
\hline
$\chi^2|_{\rm XENON1T}$ & 167.08 & 167.69 & 167.07 & 166.62 & 167.11 & 167.11 \\
$\chi^2|_{\rm XENONnT}$ & 19.98 & 20.05 & 20.92 & 20.13 & 19.95 & 19.95 \\
$\chi^2|_{{\rm XENONnT}_{\rm 20\,ton-yr}}$ & {\bf 7.54} & {\bf 12.62} & {\bf 32.01} & {\bf 14.35} & 0 & 0 \\
$N_{\rm SK}$ & 0 & 0 & 4.76 & 54.60 & 26.77 & 29.52 \\
$N_{\rm HK}$ & 0 & 0 & 109 & {\bf 1261} & {\bf 618} & {\bf 682} \\
\hline
\hline
\end{tabular}
%\end{ruledtabular}
\end{adjustbox}
\label{2HDM}
\end{table}

The main contributions to the FB energy are the  
Fermi gas kinetic energy, Yukawa potential energy, and the potential energy difference between the false and true vacua~\cite{Marfatia:2021hcp}.
The magnitude of the Yukawa energy increases as the FB temperature decreases and the
the interaction length $L_\phi$ increases:
\begin{eqnarray}
\label{eq:L_phi}
     L_\phi(T)\equiv \frac{1}{m_\phi(T)} \equiv \left(\frac{d^2 V_\mathrm{eff}}{d\phi^2}\bigg|_{\phi=0}\right)^{-1/2}=(2D(T^2-T^2_0))^{-1/2} \,.
    \end{eqnarray}
Eventually, at temperature $T_\phi$, $L_\phi$ equals the mean distance between $\chi$'s in the FB, i.e., $L_\phi \simeq R_{\rm FB}/Q^{1/3}_{\rm FB}$, and the (negative) Yukawa energy becomes comparable in magnitude to the total FB energy.
When no stable solution for the FB radius can be found, the FB collapses to a Schwarzschild PBH with mass given by the FB mass at $T_\phi$: $M_{\rm PBH}=M_{\rm FB}(T_\phi)$. The PBH mass function is monochromatic because the FB mass function is. Bounds on the PBH fraction of the energy density of the Universe when they form are often placed in terms of~\cite{Carr:2020gox}
\begin{equation}
\beta'\equiv \gamma^{1/2} \left(\frac{g_*(T_\phi)}{106.75} \right)^{1/4} \frac{\rho_{\rm PBH}(T_\phi)}{\rho(T_\phi)}\,,
\end{equation}
where $\gamma$ is the ratio of the PBH mass to the Hubble horizon mass in the radiation dominated epoch, and we have taken the Hubble constant to be $1.44\times 10^{-42}$~GeV.  For our production mechanism~\cite{Marfatia:2021hcp},
\begin{eqnarray}
\beta' & \equiv & 4.58\times 10^{-12}{T_\phi \over {\rm MeV}} \left({M_{\rm PBH}\over 10^{-18} M_\odot}\right)^{1/2}\,
\frac{\rho_{\rm PBH}(T_{\phi})}{\rho(T_{\phi})}\,.
\label{beta'}
\end{eqnarray}

During FB formation, $\phi$ can be copiously produced via $\chi \bar{\chi}\to \phi \phi$ with number density $2n_{\bar\chi}(T_\star)$. Since $m_\phi \sim T_\star$ for $D\sim 1$, $\phi$ is nonrelativistic, and its number density will evolve like matter until today.
The relic density of $\phi$ may very well overclose the Universe. 
To avoid this, we allow $\phi$ to decay to a pair of massless dark scalars $\mathcal{S}$, which preserves $g^{\rm D}_\ast=4.5$ and contributes to the effective number of extra neutrinos $\Delta N_{\rm eff}$.
We require $\Delta N_{\rm eff}\leq 0.5$~\cite{Marfatia:2021twj}.
A trilinear coupling $\mu \phi \mathcal{SS}$ in the Lagrangian makes $\phi$ decay with a lifetime~\cite{Matsumoto:2018acr},
\begin{eqnarray}
\tau_{\phi \to \mathcal{SS}}\simeq 0.66\times 10^{-19} \left( \frac{\rm MeV}{\mu} \right)^2
\left( \frac{m_\phi}{\rm MeV} \right)~[{\rm sec}]\,.
\end{eqnarray}
Setting e.g., $\mu=0.01~{\rm MeV}$, ensures that $\phi$ does not overclose the Universe.
On replacing $\mathcal{S}$ by its vacuum expectation value, the trilinear coupling modifies the parameters in $V_{\rm eff}$ (Eq.~\ref{eq:quartic_potential})~\cite{Beniwal:2018hyi,DiBari:2021dri}; the term linear in $\phi$ can be removed by a translation of the field so that the false vacuum lies at $\langle \phi \rangle=0$. However, since the parameters are free, they can be interpreted as the values corrected by the new coupling.

\bigskip

\section{$\chi$ flux from PBH evaporation}
\label{sec:PBH_evaporation}

The dark fermion $\chi$ will be emitted by a PBH if $m_\chi(T_\star)$ is smaller than the Hawking temperature $T_{\rm PBH}$, which describes the thermal spectrum from PBH evaporation:
    \begin{eqnarray}
    T_{\rm PBH}\simeq 5.3~{\rm MeV}\times \left( \frac{10^{-18}M_\odot}{M_{\rm PBH}} \right)\,.
    \end{eqnarray}
    However, even if  $m_\chi(T_\star) > T_{\rm PBH}$, some $\chi$ will be emitted because of the long tail of the approximately black-body spectrum.
    The differential spectrum of $\chi$ is given by~\cite{Hawking:1974rv,Hawking:1975vcx}
    \begin{eqnarray}
    \frac{dN_\chi}{d\mathcal T dt}=
    \frac{2 \Gamma_\chi(\mathcal T,M_{\rm PBH})}{\pi (e^{(\mathcal T+m_\chi)/T_{\rm PBH}}+ 1)}\,,
    \end{eqnarray}
    where $\mathcal T$ is the kinetic energy of $\chi$,
    and $\Gamma_\chi$ is the graybody factor which characterizes deviations from a black-body spectrum due to the nonzero probability of $\chi$ to fall back into the 
    gravitational well of the PBH.
        We utilize the BlackHawk v2.1~\cite{Arbey:2019mbc,Arbey:2021mbl} package
    to calculate the spectrum from PBH evaporation.
    
    The $\chi$ flux on Earth has a component from 
    extragalactic PBHs and from PBHs in the Milky Way. 
   We compute both contributions, 
   although the former is more robust as it does not rely on the DM density profile 
   near the galactic center.
    The isotropic extragalactic flux is given by
    \begin{eqnarray}
    \frac{d\Phi^{\rm EG}}{d\mathcal T}=\int^{\mathrm{min}(t_{\mathrm{eva}},t_0)}_{t_\phi}{c\,dt[1+z(t)]\frac{f_\mathrm{PBH}\rho_{\mathrm{DM}}}{M_\mathrm{PBH}}\frac{d^2N_\chi}{d\mathcal Tdt}\bigg|_{\tilde{E}=\sqrt{(E^2-m_\chi^2)(1+z(t))^2+m_\chi^2}}}\,,
    \end{eqnarray}
where $f_{\rm PBH}$ is the fraction of DM in the form of PBHs,  and $\rho_\mathrm{DM}=1.27\mkern5mu \mathrm{GeV }/\mathrm{m}^{3}$ is the average DM density of the Universe in the current epoch~\cite{Planck:2018vyg}. For PBHs that have not survived until today, $f_{\rm PBH}$ is interpreted 
as the value today had they not evaporated away. 
In the integration limits, $t_\phi$ is the time of PBH formation (corresponding to $T_\phi$), $t_\mathrm{eva}$ is the lifetime of the PBH, and $t_0=13.77\times10^9\mkern5mu\mathrm{yr}$ is the age of the Universe. Note that $f_{\rm PBH}$  and $\beta'$ are related by~\cite{Carr:2020gox}
\begin{equation}
f_{\rm PBH}=3.78\times 10^{17} \beta'  \left({10^{-18} M_\odot \over M_{\rm PBH}}\right)^{1/2}\,.
\label{fbeta}
\end{equation}

The time-redshift relation is obtained by numerically solving
$dt = -dz/[(1+z) H(z)]$
where $H^2(z) = H_0^2 [\Omega_m(1+z)^3 + \Omega_r(1+z)^4 + (1-\Omega_m-\Omega_r)]$, with
$\Omega_m=0.315$, $\Omega_r=9.23\times 10^{-5}$, and $H_0=67.4~{\rm km/s/Mpc}$~\cite{Planck:2018vyg}.
The $\Delta N_{\rm eff}$ contribution to $\Omega_r$ by the FOPT can be neglected since the $\chi$ flux emitted before matter-radiation equality at $z\simeq 3400$, when  
$t\simeq 5.1\times 10^4$~year, is sufficiently redshifted to be undetectable.

The galactic flux is given by
\begin{eqnarray}
\frac{d\Phi^{\rm MW}}{d\mathcal T}= \int \frac{d\Omega}{4 \pi}
\int_{0}^{50\,\rm{kpc}}  d\ell \frac{f_{\rm PBH}\, \rho_{\rm MW}(r(\ell,\psi))}{M_{\rm PBH}} \frac{dN_\chi}{d\mathcal{T} dt}\bigg|_{t=t_0}\,,
\end{eqnarray}
where 
\begin{eqnarray}
&& r(\ell, \psi)=\sqrt{r^2_\odot -2\ell r_\odot \cos \psi + \ell^2}\,, \nonumber 
\end{eqnarray}
is the galactocentric distance of the PBH, with $\ell$ the line-of-sight distance to the PBH, 
$r_\odot=8.5~{\rm kpc}$ the distance of the Sun from the galactic center, and $\psi$ the angle between them.
We take the DM density of the Milky Way to follow the Navarro-Frenk-White profile~\cite{Navarro:1996gj},
\begin{eqnarray}
&& \rho_{\rm MW}(r)=\rho_\odot \left[ \frac{r_\odot}{r} \right]
\left[ \frac{1+r_\odot/r_s}{1+r/r_s} \right]^2\,,
\end{eqnarray}
where
$r_s=20~{\rm kpc}$ is the scale radius and $\rho_\odot=0.4~{\rm GeV/cm^{3}}$ is the local DM density. Then, the total 
$\chi$ flux is
$$
\frac{d\Phi}{d \mathcal{T}}=\frac{d\Phi^{\rm EG}}{d \mathcal{T}}+\frac{d\Phi^{\rm MW}}{d \mathcal{T}}\,.
$$

\bigskip
    
\section{$\chi-e$ scattering event rates}
\label{sec:event_rate}

We are interested in the evaporation of PBHs that produces a $\chi$ spectrum in an 
energy range from keV to GeV so that
the XENON1T/XENONnT and SK/HK detectors can detect a signal.
The $\chi+e \to \chi+e$ differential cross section per unit recoil energy $E_r$ on free electrons in the lab frame
is given by~\cite{Cao:2020bwd,Calabrese:2022rfa}
\begin{eqnarray}
\label{eq:DMe_CrossSection}
\frac{d\sigma}{dE_r}=\frac{\sigma_{\chi e}\Theta(E_{r}^{\rm max}-E_r)}{8 \mu^2_{\chi e}\tilde{p}^2}(2m_e+E_r)(2m^2_\chi+m_e E_r)\,,
\end{eqnarray}
where $\mu_{\chi e}\equiv m_\chi m_e/(m_\chi+m_e)$ is the reduced mass of the $\chi-e$ system, $\tilde{p}$ is the magnitude of the three-momentum of $\chi$, and the maximum allowed recoil energy is
\begin{eqnarray}
E_{r}^{\rm max}=\frac{2m_e \mathcal T (\mathcal T +2m_\chi)}{(m_e+m_\chi)^2+2m_e \mathcal T }\,.
\end{eqnarray}
We assume that $\sigma_{\chi e}$ is independent of $m_\chi$.  As mentioned earlier, we require
$\sigma_{\chi e} < 10^{-31}~{\rm cm^2}$ so that propagation in the atmosphere and Earth does not
attenuate the $\chi$ flux.

\bigskip

\subsection{XENON1T and XENONnT}

At the XENON1T/XENONnT detectors, a boosted $\chi$ from PBH evaporation can ionize the Xe atom via $\chi+{\rm Xe} \to \chi + {\rm Xe}^*+e^-$,
where $\chi$ scatters with a bound electron in the Xe atom and produces an electron recoil signal. Another source of
such a signal is the $\beta$-decay of trace amounts of tritium in xenon.
The excess in electronic-recoil events between $2-3$~keV in XENON1T data was suspected to be a result of tritium contamination~\cite{XENON:2020rca}.
This has been confirmed by the XENONnT experiment 
by reducing the tritium rate to below 15 events/ton-yr at 90\% CL  and the
 background to five times lower than in XENON1T.
The XENON1T excess is excluded by XENONnT at $4\sigma$~\cite{Aprile:2022vux}.
We use the combined data from the 0.65 ton-year and 1.16 ton-year exposures with recoil energy ranges $[1~{\rm keV},210~{\rm keV}]$ and $[1~{\rm keV},30~{\rm keV}]$ 
from XENON1T~\cite{XENON:2020rca} and XENONnT~\cite{Aprile:2022vux}, respectively. Note that the XENON1T and XENONnT data are complementary because the XENONnT data are 
an order magnitude more sensitive than XENON1T data between $[1~{\rm keV},30~{\rm keV}]$, 
while the XENON1T spectrum extends to higher recoil energy. 

We estimate the future sensitivity of XENONnT, which is expected to accumulate 20 ton-year of data, by assuming that statistical uncertainties dominate and scaling the current uncertainty by $\sqrt{1.16/20}$~\cite{XENON:2020kmp}. %

The differential event rate at XENON1T/XENONnT is given by
\begin{eqnarray}
\label{eq:EventRate}
\frac{dR}{dE_r}=n_t\,\tilde{F}(E_r) \int d\mathcal T \frac{d\Phi}{d\mathcal T} \sum_{n,l} \frac{d\sigma^{n,l}}{dE_r}\,,
\end{eqnarray}
where $n_t=4.59\times 10^{27}$ is the number of Xe atoms per ton, and the Fermi factor $\tilde{F}(E_r)$ describes the distortion of the final-state electron wavefunction by the atom. Here, $d\sigma^{n,l}/dE_r$ is the differential
cross section for scattering of $\chi$ on a bound electron with principal quantum number $n$ and orbital quantum number $l$. Because the binding energy is non-negligible compared to the energy of $\chi$, the ionization process is not trivial. For details of how this is calculated see Ref.~\cite{Calabrese:2022rfa}, which uses the results of Refs.~\cite{Kopp:2009et,Lee:2015qva,Catena:2019gfa}. We apply a Gaussian smearing to the spectrum in Eq.~(\ref{eq:EventRate}) to model the detector resolution~\cite{XENON:2020rca}, and correct it with the detector efficiency~\cite{XENON:2020rca,Aprile:2022vux}. 

To find the parameter space excluded by current XENON1T/XENONnT data and to evaluate the sensitivity of future
XENONnT data, we define in obvious notation on the smeared distributions,
\begin{eqnarray}
\label{eq:chi_sqaure}
\chi^2\equiv \sum_i \left(  \frac{ \left. \frac{dR}{dE_r}\right|_i + \left. \frac{dR_{\rm bkgd}}{dE_r}\right|_i \left.  - \frac{dR_{\rm obs}}{dE_r}\right|_i}{\sigma_i} \right)^2\,.
\end{eqnarray}
The uncertainty and background in each recoil energy bin $i$ is taken from Refs.~\cite{XENON:2020rca,Aprile:2022vux}.
A point in the parameter space is excluded at $2\sigma$ if
$\Delta \chi^2 = \chi^2-\chi^2_{\rm bkgd}\geq 4$. 
For the background only hypothesis and current data, $\chi^2_{\rm bkgd}|_{\rm XENON1T}=167.1$ and $\chi^2_{\rm bkgd}|_{\rm XENONnT}=19.95$.
We assume that future 20 ton-year XENONnT data are consistent with the background expectation, and thus $\chi^2_{\rm bkgd}|_{{\rm XENONnT}_{\rm 20\,ton-yr}}=0$.

\bigskip

\subsection{Super-Kamiokande and Hyper-Kamiokande}

$\chi-e$ scattering in the SK and HK detectors will produce Cherenkov radiation.
The 161.9 kiloton-year SK dataset has $N^{\rm SK}_{\rm obs}=4042$ electron events in the recoil energy range $[0.1~{\rm GeV},1.33~{\rm GeV}]$, which is compatible with the background expectation, $N^{\rm SK}_{\rm bkgd}=3992.9$~\cite{Super-Kamiokande:2017dch}.
Since the electron binding energies in hydrogen and oxygen are much smaller that in xenon, it is reasonable to treat the 
target electrons as free and at rest.
Then, the differential event rate is computed by inserting Eq.~(\ref{eq:DMe_CrossSection}) into an equation of the form of Eq.~(\ref{eq:EventRate}) with $n_t=3.34\times 10^{28}$, the Fermi factor removed, and multiplied by a 93\% detector efficiency. The event number at SK is
\begin{eqnarray}
N^{\rm SK}_{\rm PBH}= 161.9~[{\rm kton-yr}]\times \int^{1.33~{\rm GeV}}_{0.1~{\rm GeV}} dE_r \frac{dR}{dE_r}\,.
\end{eqnarray}
We define $2\sigma$ exclusion by $N^{\rm SK}_{\rm PBH}/\sqrt{N^{\rm SK}_{\rm PBH}+N^{\rm SK}_{\rm bkgd}}\geq 2$, which corresponds to $N^{\rm SK}_{\rm PBH} = 129$.
In the future, HK is expected to collect a dataset with a 3.74 Mton-year exposure~\cite{Hyper-Kamiokande:2018ofw}.
To forecast the sensitivity of the HK dataset, we follow the above procedure with the background event number appropriately rescaled. Then, a $2\sigma$ signal occurs for $N^{\rm HK}_{\rm PBH} = 610$.

In Fig.~\ref{fig:scan}, the yellow points are excluded by current 
XENON1T/XENONnT/SK data, 
and the red points will be probed by future XENONnT/HK data.

\bigskip

\section{Correlated signals}
\label{sec:scan}

\begin{figure}[t!]
\centering
\includegraphics[height=1.5in,angle=0]{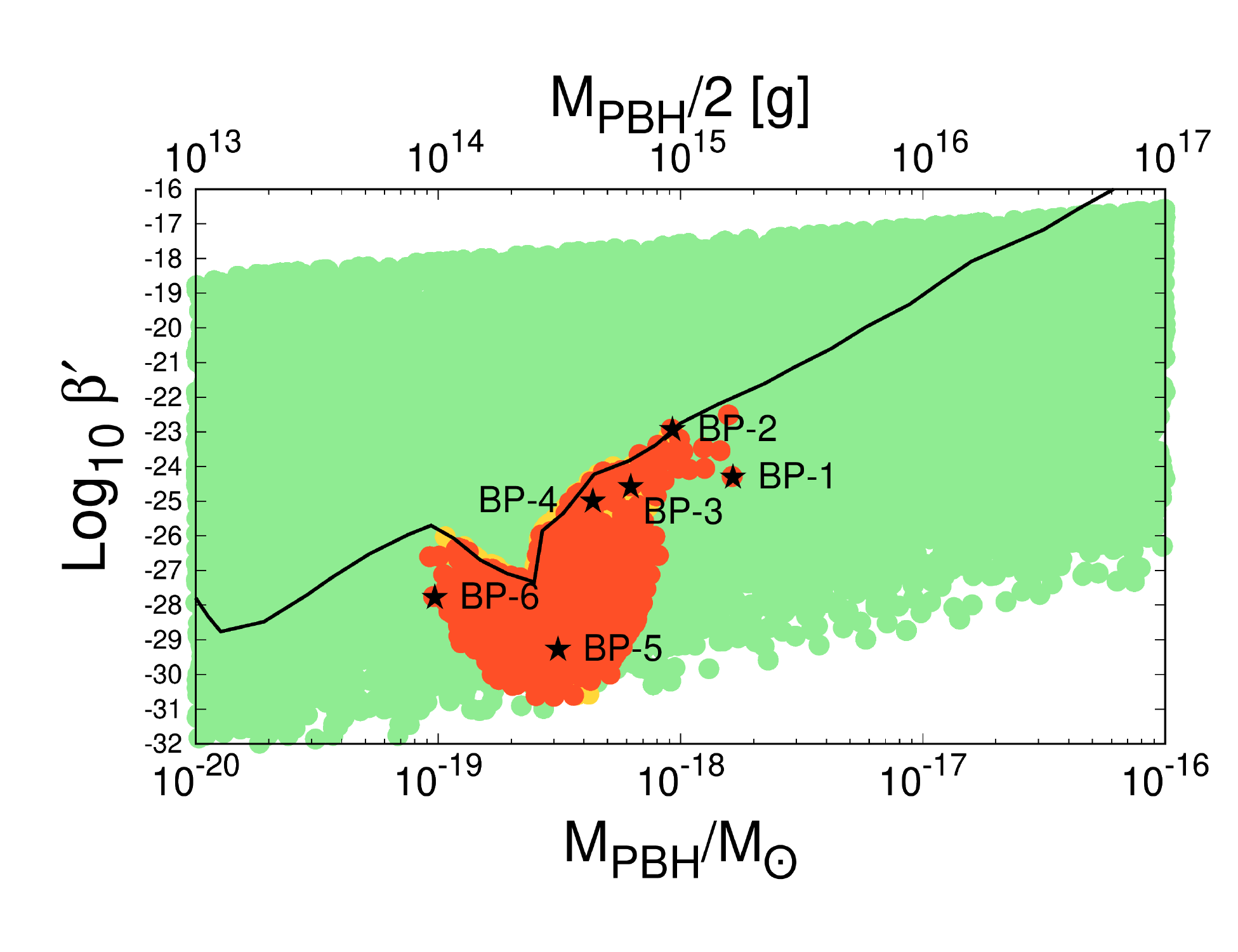}
\includegraphics[height=1.5in,angle=0]{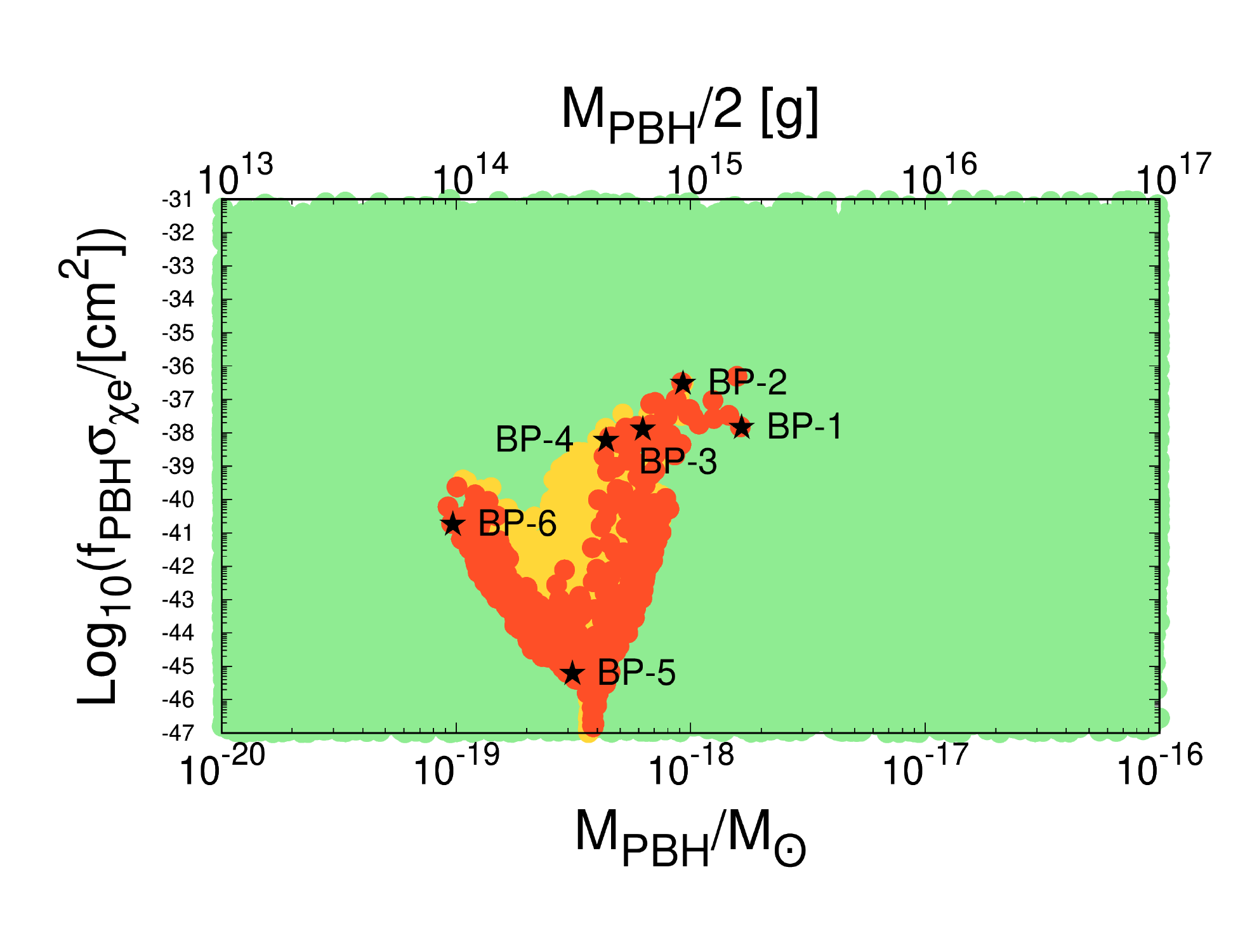}
\includegraphics[height=1.5in,angle=0]{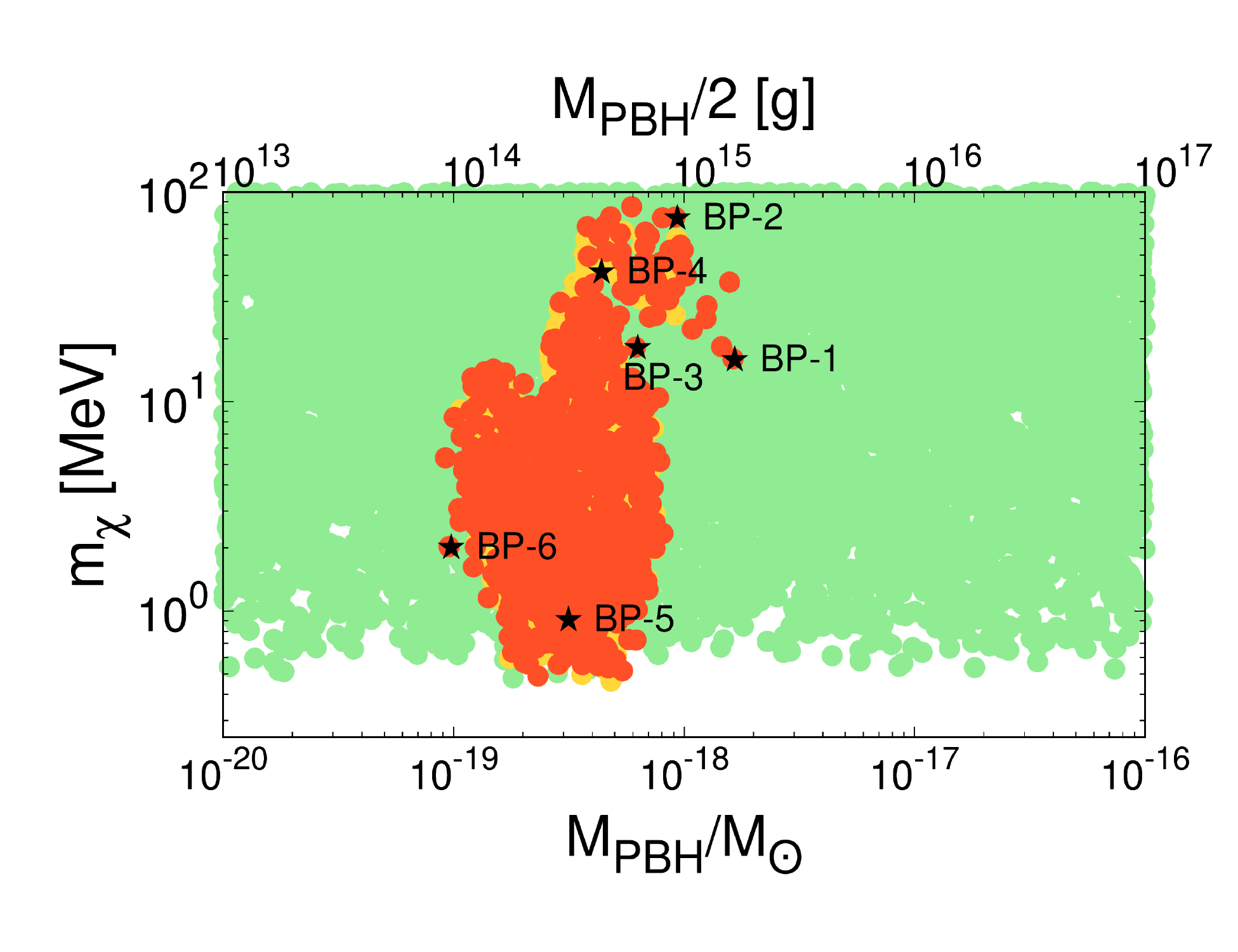}
\includegraphics[height=1.5in,angle=0]{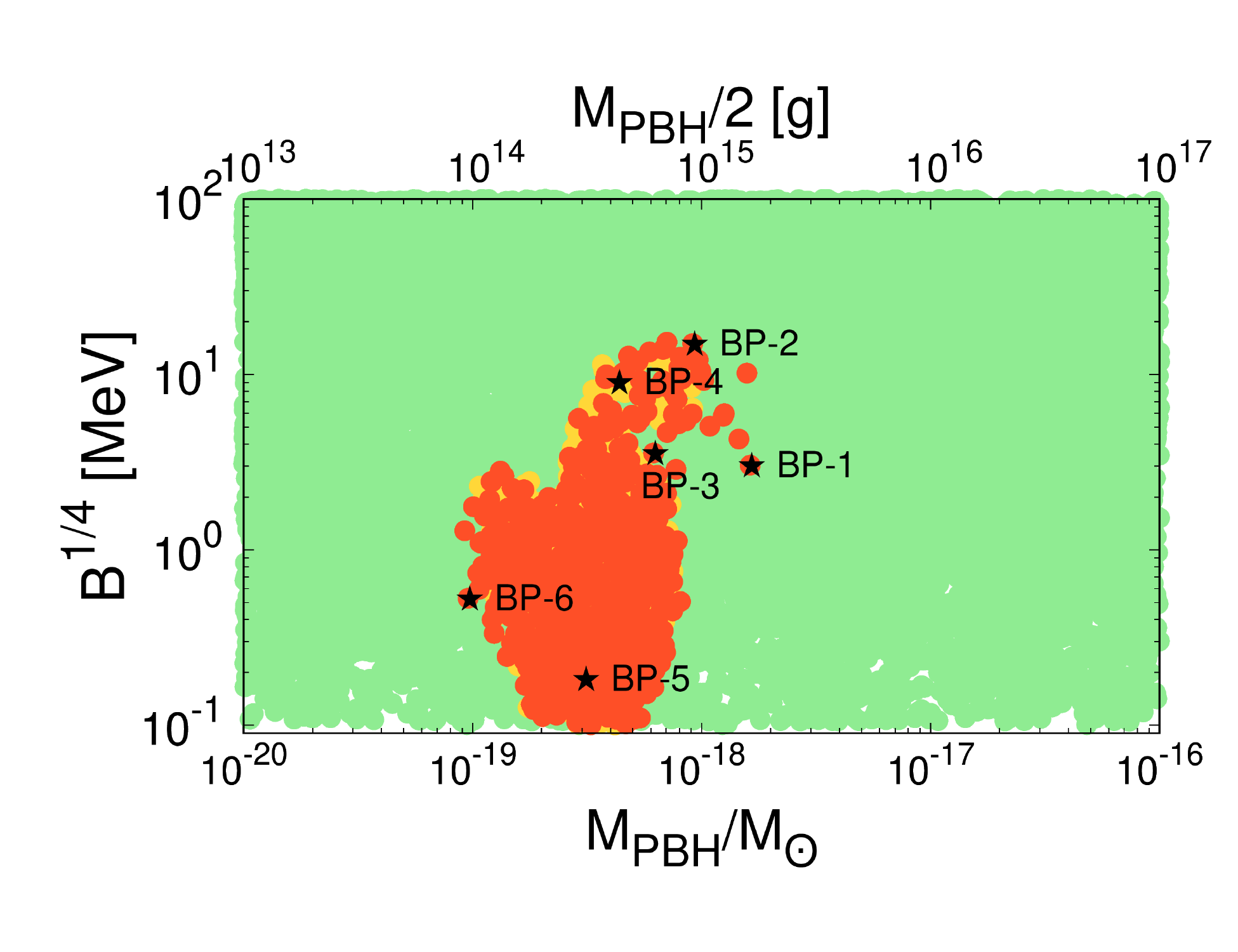}
\includegraphics[height=1.5in,angle=0]{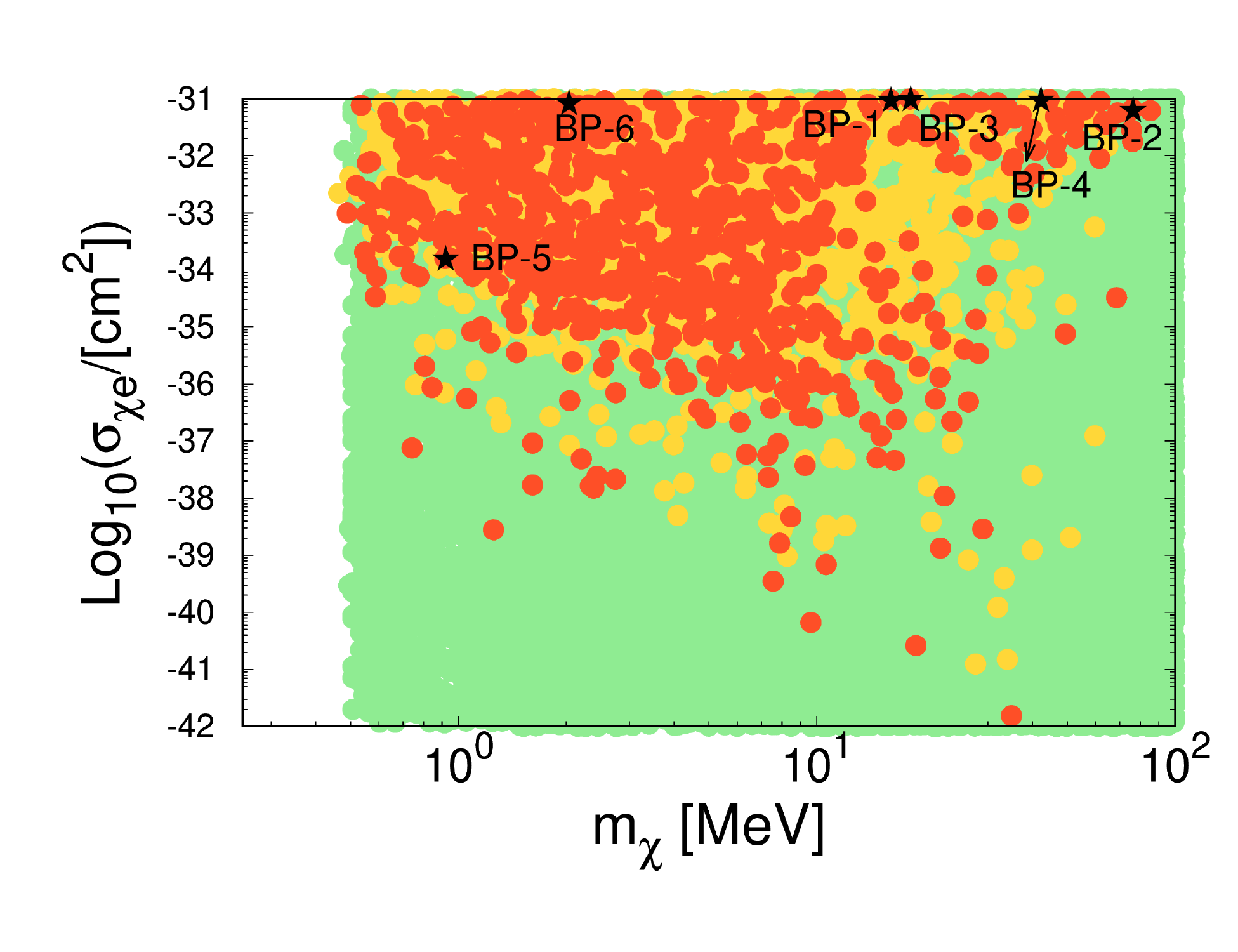}
\includegraphics[height=1.5in,angle=0]{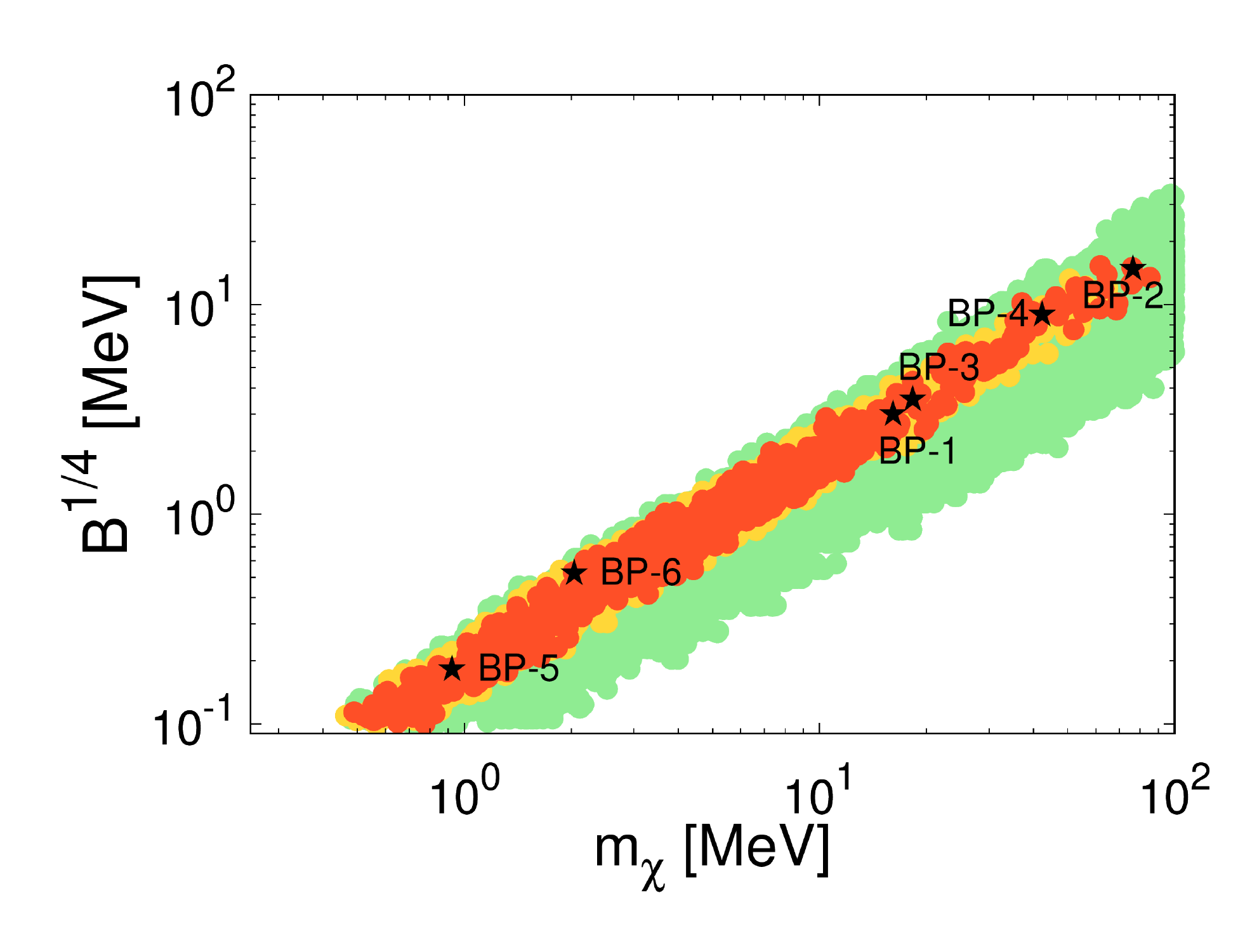}
\includegraphics[height=1.5in,angle=0]{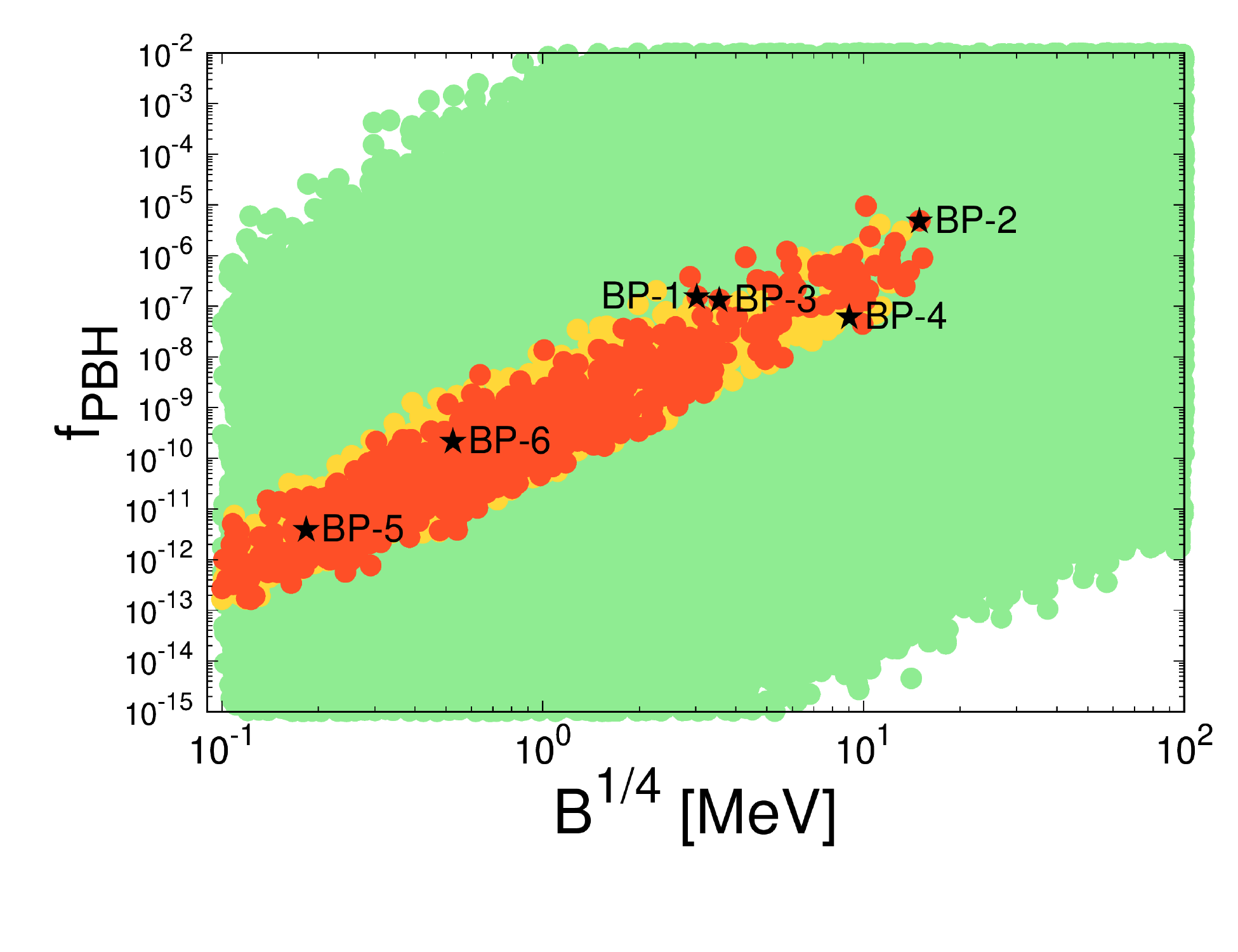}
\includegraphics[height=1.5in,angle=0]{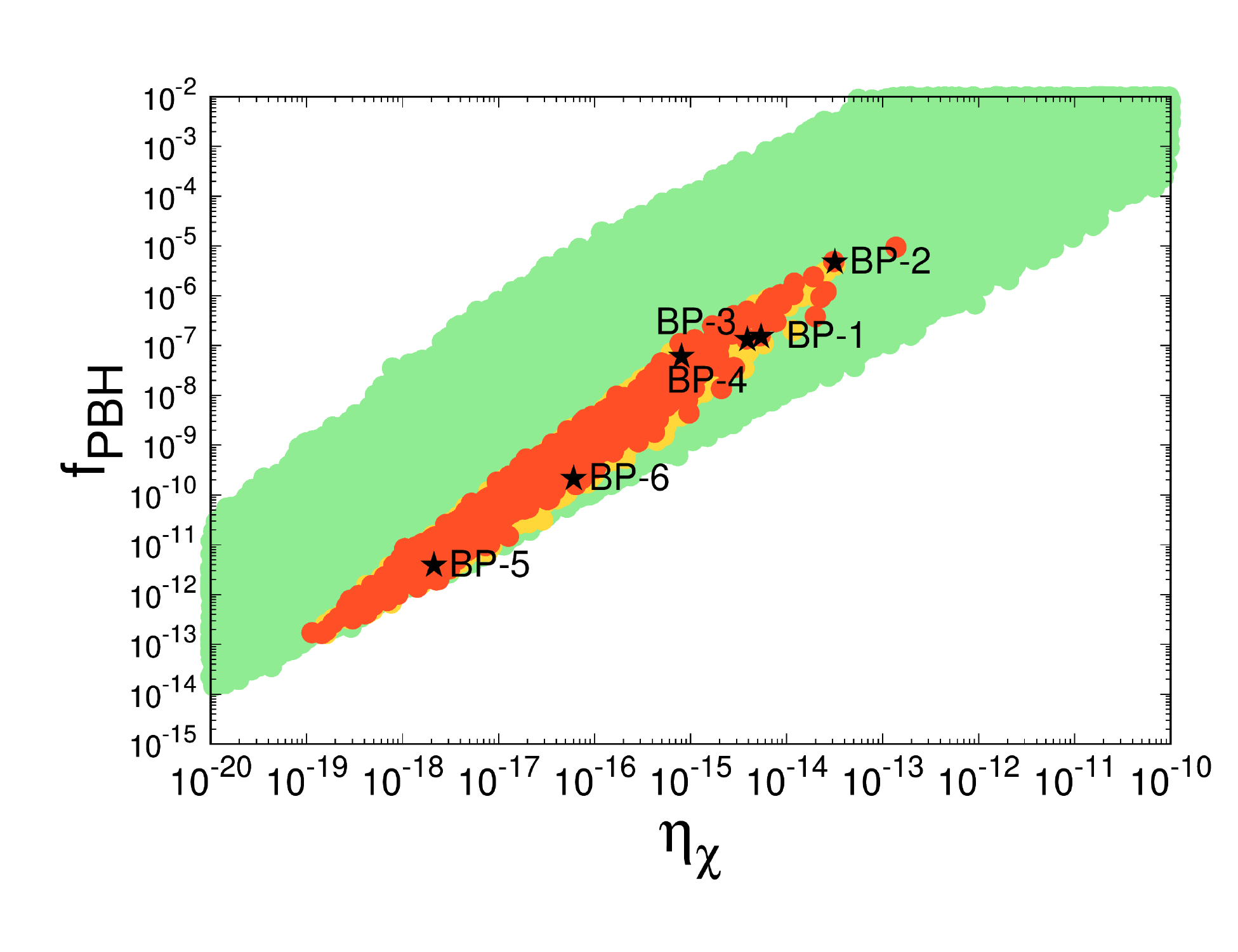}
\caption{\small \label{fig:scan}
The parameter space in which a GW signal can be detected at THEIA/$\mu$Ares is shown with green points. 
The yellow points are excluded by current 
XENON1T/XENONnT/SK data at $2\sigma$, and the red points are consistent with these data at $2\sigma$, but can produce a $2\sigma$ DM signal in future XENONnT/HK data and a GW signal at THEIA/$\mu$Ares.
The solid black curve in the top left panel shows the current bound from observations of the extragalactic 
gamma-ray background and from the damping of small scale CMB anisotropies. The stars mark the six benchmark points in Table~\ref{tab:BP}. All points satisfy $\Omega_{\rm PBH} h^2 \leq 0.12$, $\Delta N_{\rm eff}\leq 0.5$ and $m_\chi(T_\star)-m > 2 T_{{\rm SM}\star}$.
}
\end{figure}

We fix $A=0.1$ and scan the other parameters of the effective potential in Eq.~(\ref{eq:quartic_potential}), 
the temperature ratio of the dark and SM sectors $T_\star/T_{\rm SM\star}$ during the phase transition,
the Yukawa coupling $g_\chi$, the bare mass $m$, and $\sigma_{\chi e}$  in the ranges,
\begin{eqnarray}
0.05 & \leq & \lambda \leq 0.2\,, \ \ \
0.1  \leq  B^{1/4}/{\rm MeV} \leq  10^4\,, \ \ \ 
0.01  \leq  C/{\rm MeV} \leq  10^4\,, \nonumber \\
0.1 & \leq & D \leq 10\,, \ \ \
0.3  \leq  T_\star/T_{\rm SM\star} \leq 1.0\,, \ \ 
0.01  \leq  g_\chi \leq \sqrt{4 \pi}\,, \nonumber \\
10^{-3} & \leq & m/B^{1/4}  \leq 10\,, \ \ \
10^{-42}  \leq  \sigma_{\chi e}/{\rm cm^2} \leq 10^{-31}\,.
\end{eqnarray}
 The value of $\eta_\chi$ is adjusted
to ensure that $\Omega_{\rm PBH}h^2 \leq \Omega_{\rm DM}h^2\simeq 0.12$. 

The result of the scan is shown in Fig.~\ref{fig:scan}. All points 
satisfy $\Omega_{\rm PBH} h^2 \leq 0.12$ and $\Delta N_{\rm eff}\leq 0.5$. For the green points the stochastic GW background can be observed 
by the proposed THEIA~\cite{THEIA} and $\mu$Ares~\cite{Sesana:2019vho} telescopes.
In the region defined by the red points, a GW signal can detected at THEIA/$\mu$ARES, and PBH evaporation produces a DM flux that can be detected in future XENONnT and HyperK data at the $2\sigma$ CL.
 All red points are consistent
with current XENON1T/XENONnT~\cite{XENON:2020rca,Aprile:2022vux}  and 
SK~\cite{Super-Kamiokande:2017dch} data at $2\sigma$.
The region above the solid black curve in the $(M_{\rm PBH}/M_\odot,\beta')$ plane is excluded by observations of the extragalactic $\gamma$-ray background and by the damping of small scale CMB anisotropies by PBH evaporation after
recombination~\cite{Carr:2020gox}.
The yellow points are excluded by current XENON1T/XENONnT and SK data at $2\sigma$. Interestingly, many of them lie below the
solid black line, thus showing that these data are more constraining than $\gamma$-ray/CMB data in a part of  the parameter space.
Using Eq.~(\ref{fbeta}), we display how $f_{\rm PBH}$ relates to $B^{1/4}$ and $\eta_\chi$.

Because SK has a higher threshold energy and a larger fiducial volume than XENON1T and XENONnT, it has greater sensitivity for $M_{\rm PBH}/M_\odot \lesssim 10^{-18}$. On the other hand, XENON1T/XENONnT is more sensitive to heavier PBHs (and lower Hawking temperatures) due to its lower energy threshold. 
The falling edge of the red region above $M_{\rm PBH}/M_\odot \sim 5\times 10^{-19}$ in the $(M_{\rm PBH}/M_\odot,m_\chi)$ plane is a result of lighter PBHs emitting
heavier $\chi$. This also corresponds to PBHs with lifetimes longer than the age of the Universe.  

\begin{figure}[t!]
\centering
\includegraphics[height=4.2in,angle=270]{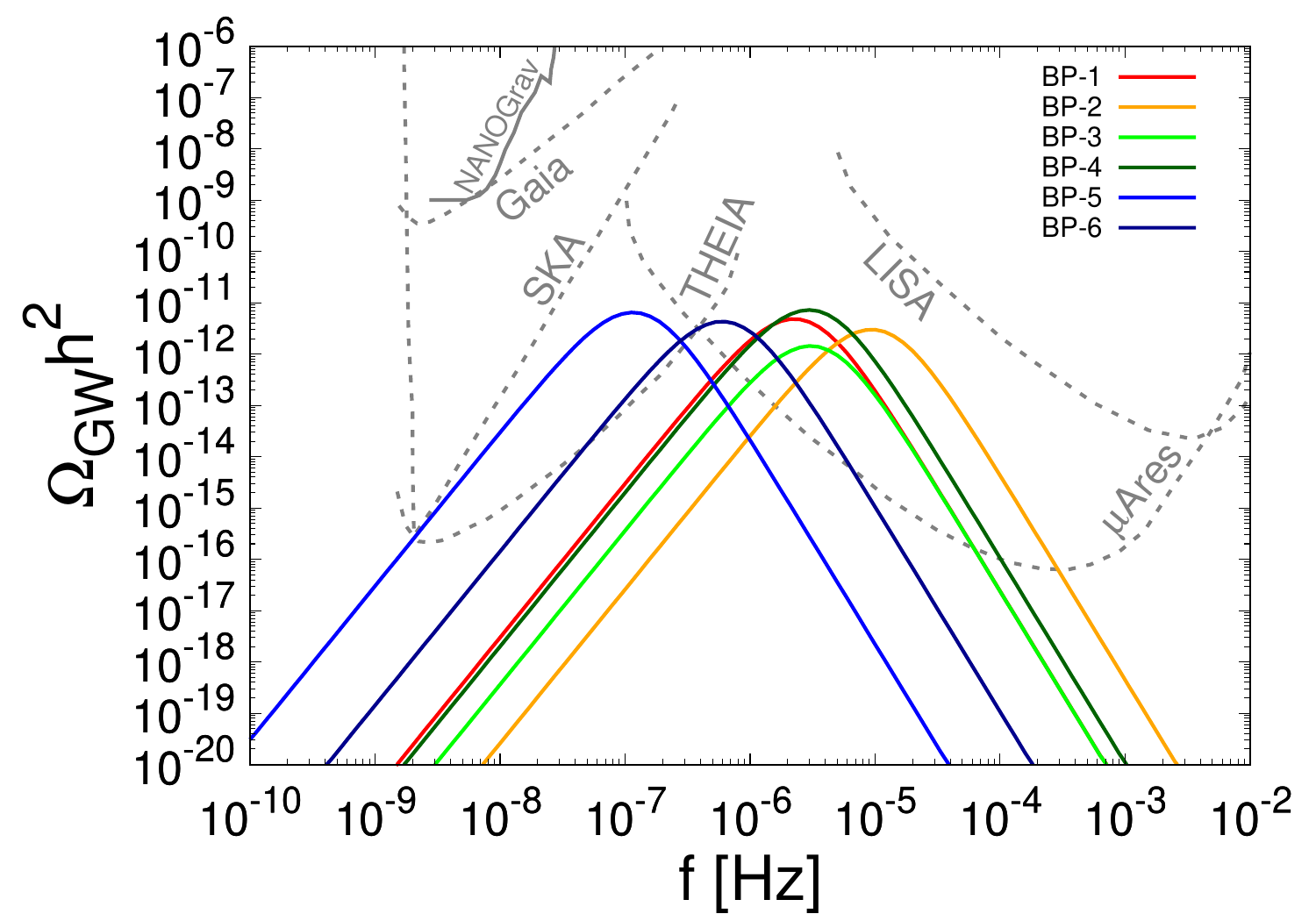}
\caption{\small \label{fig:GW}
Gravitational wave spectra for the benchmark points in Table~\ref{tab:BP}.
}
\end{figure}

The six benchmark points (BPs) listed in Table~\ref{tab:BP} are selected from the red regions in Fig.~\ref{fig:scan}.
Figure~\ref{fig:GW} shows their GW spectra, obtained using the procedure of Ref.~\cite{Marfatia:2020bcs}.
{\bf BP-1}, {\bf BP-2}, and {\bf BP-3}, with $M_{\rm PBH}/M_\odot \gtrsim 5\times 10^{-19}$ and $T_{\rm PBH} \lsim 10.6$~MeV, produce mostly
nonrelativistic $\chi$ (since $m_\chi(T_\star) > T_{\rm PBH}$), and can only be probed by 20 ton-yr of XENONnT data because of its low energy threshold. 
 However, although {\bf BP-4} produces nonrelativistic $\chi$, future XENONnT and HK data will be sensitive to {\bf BP-4} at $2\sigma$ because $\chi$ is relatively heavy.
{\bf BP-5}, and {\bf BP-6} can be detected at HK despite their small $f_{\rm PBH} \sigma_{\chi e}$ values due to HK's large exposure.

\section{Summary}
\label{sec:summary}

It has been recognized that primordial black holes may form during a first-order phase transition. It has also been recognized that boosted dark matter can arise from PBH evaporation. We have connected these two mechanisms in
a predictive framework and shown that the dark sector particles $\chi$ that form Fermi balls during a FOPT, which then collapse to PBHs, can be emitted as DM in PBH evaporation.
If $\chi$ couples to electrons, this DM flux can be detected at XENONnT and Hyper-Kamiokande. The signal has a gravitational wave counterpart at THEIA and $\mu$Ares. CMB-S4 will measure a nonzero value of $\Delta N_{\rm eff}$.

\section*{Acknowledgements}  
D.M. is supported in
part by the U.S. DOE under Grant No. de-sc0010504. 
P.T. is supported in part by the Ministry of Sciences and Technology under
grant number MoST-111-2112-M-007-012-MY3.

%%%%%%%%%%%%%%%%%%%%%-------------------

\end{document}